# Optical Antenna-based Fluorescence Correlation Spectroscopy to Probe the Nanoscale Dynamics of Biological Membranes

*Pamina M. Winkler[†], Raju Regmi[‡,†], Valentin Flauraud[§], Jürgen Brugger[§], Hervé Rigneault[‡], Jérôme Wenger[\*‡], María F. García-Parajo[\*†⊥]*

[†] ICFO-Institut de Ciencies Fotoniques, The Barcelona Institute of Science and Technology, Barcelona, Spain; [‡] Aix Marseille Univ, CNRS, Centrale Marseille, Institut Fresnel, Marseille, France; [§] Microsystems Laboratory, Institute of Microengineering, Ecole Polytechnique Fédérale de Lausanne, 1015 Lausanne, Switzerland; [⊥] ICREA, Pg. Lluís Companys 23, 08010 Barcelona, Spain.

\*E-mail: maria.garcia-parajo@icfo.eu; jerome.wenger@fresnel.fr

The plasma membrane of living cells is compartmentalized at multiple spatial scales ranging from the nano- to the meso-scale. This non-random organization is crucial for a large number of cellular functions. At the nanoscale, cell membranes organize into dynamic nano-assemblies enriched by cholesterol, sphingolipids and certain types of proteins. Investigating these nano-assemblies known as lipid rafts is of paramount interest in fundamental cell biology. However, this goal requires simultaneous nanometer spatial precision and microsecond temporal resolution which is beyond the reach of common microscopes. Optical antennas based on metallic nanostructures efficiently enhance and confine light into nanometer dimensions, breaching the diffraction limit of light. In this Perspective, we discuss recent progress combining optical antennas with fluorescence correlation spectroscopy (FCS) to monitor microsecond dynamics at nanoscale spatial dimensions. These new developments offer numerous opportunities to investigate lipid and protein dynamics in both mimetic and native biological membranes.

The plasma membrane is a complex, versatile and essential signaling interface that separates the cell cytoplasm from the extracellular space.[1,2] Its spatiotemporal organization and biological function are intricately interlaced at the nanoscale.[3–5] The heterogeneous landscape of the cell membrane is shaped by a variety of lipids and proteins that differ in their physico-chemical properties. Sphingolipids, cholesterol and certain types of proteins such as glycosylphospatidylinositol-anchored proteins (GPI-APs) can assemble into dynamic nanoscale clusters or nanodomains, also known as lipid rafts.[6,7]. Although lipid rafts have been implicated in a large number of cellular functions[5,8–15], their existence in living cells have been heavily debated given the enormous challenge associated with observing these transient nanodomains.[7,16–23] This Perspective aims to briefly review and to place into a broader context very recent works using advanced nano-optical techniques to investigate lipid rafts and their dynamics in biological membranes.

The basis for understanding cell membrane structure was proposed nearly 50 years ago by Singer and Nicolson.[24] This fluid mosaic model captures the general characteristics of the cell membrane as a lipid bilayer dressed with embedded proteins. However, intensive research in the last twenty years has revealed that biological membranes are highly heterogeneous and with a much higher complex architecture that goes well beyond what it was initially proposed by the fluid mosaic model. Within the plane of the membrane, certain types of proteins, sphingolipids and cholesterol arrange in transient nanoscopic domains, also denoted as lipid rafts.[1,2,6,17,21] These highly dynamic and fluctuating nanoscale assemblies can be stabilized in the presence of lipid- or protein-mediated activation events to compartmentalize cellular processes.[2,18] By means of physically segregating specific molecular components within the membrane, lipid rafts are believed to modulate the activity of raft-associated proteins, and influence signaling and function of a broad range of membrane receptors.[5,8,10,15] Moreover, recent research indicates that the biophysical properties of lipid rafts (size, composition and dynamics) can be modulated by the proximal actin cytoskeleton[7,25,26] and components of the extracellular matrix[27–31], adding an extra complexity to the sub-compartmentalization of the plasma membrane. While the overwhelming diversity of membrane nanodomains makes their study particularly challenging, understanding the fundamental mechanisms that lead to raft formation as the first organizing principle of the cell membrane, is of paramount importance.

Artificial lipid bilayers have been extensively used as model systems since they recapitulate some of the most important features of biological membranes.[32–35] On the microscopic scale, ternary lipid membranes composed of unsaturated phospholipids, saturated sphingolipids and cholesterol separate into two distinct liquid phases which can be resolved by diffraction-limited optics: a liquid disordered (Ld) phase comprised mainly of unsaturated phospholipids and a liquid ordered (Lo) phase mostly composed of saturated lipids and cholesterol.[2,5,36] This Lo phase has been considered to represent the potential physical model for living rafts in cellular membranes.[2,5,32–35] Microscopic and stable liquid-liquid phase separation has been observed on both supported lipid bilayers (SLB) and giant unilamellar vesicles prepared from cell membrane lipid extracts.[37,38] However, such phase coexistence has remained so far largely unresolved on biological membranes. Interestingly, some studies have shown that the cell membrane in all its complexity is fully capable to phase segregate into a micrometer-sized two-phase fluid-fluid system, upon a temperature decrease[39], or through ganglioside GM1 (a raft lipid) tightening by its ligand cholera toxin-β (CTxB) at physiological temperatures[40], provided that the membrane is separated from the influence of the cortical cytoskeleton. Based on these results, it has been proposed that an underlying selective connectivity mediated by cholesterol must exist among membrane rafts even at the resting state.[2,40] This connectivity will thus be responsible for the large-scale phase segregation induced far beyond the valency of initial GM1 tightening through CTxB.[40,41] Yet, most of the experimental proof for such raft connectivity has been based on the visualization of the end stage of an activated condition and in the absence of the cytoskeleton and/or membrane traffic, where the transient rafts are amplified to coalesce into larger, stable micrometer-sized raft domains. It is only at this stage that standard fluorescence microscopy is able to observe this segregation.

In the context of fully intact living cells, early investigations on membrane organization yielded conflicting results regarding the sizes, distribution and dynamics of lipid rafts, including experimental results that refuted their existence.[2,5,16,18,23] Most of the earliest work was performed using fluorescence recovery after photobleaching (FRAP)[42,43] and more recently, using single particle tracking (SPT)[3,23,43,44] and fluorescence correlation spectroscopy (FCS).[19,23,43] FCS has been widely adopted for studying structural dynamics and biomolecular interactions on cell membranes as it features several key advantages.[19,45,46] The working principle of FCS is to analyze the temporal correlation of fluorescence intensity fluctuations.[47] This allows to determine the mean transit time averaging over thousands of single molecule diffusion events. The local molecular mobility can thus be investigated with a high temporal resolution in the sub-microsecond regime together with a broad dynamic range of timescales from microseconds to seconds.

While the FCS correlation function contains rich information on the molecular mobility, it is however hard to extract a complete description of the diffusion process (free, anomalous, constrained, directed …) out of a single FCS measurement. Alternatively, a more powerful method consists in performing diffusion measurements over a range of observation areas, as first introduced by Yechiel and Edidin in the context of FRAP.[48] This concept has been further generalized by Lenne and coworkers to establish the so-called "FCS diffusion law"[20,49], which is a graph representing the average FCS diffusion time as a function of increasing observation areas (Fig.1). Based on a series of FCS measurements for different observation areas, the shape of the FCS diffusion law allows to determine the nature of the diffusion process and the underlying membrane organization at scales smaller than the accessible experimental observation area.[20,50] Free diffusion is characterized by a strict linear proportionality between the diffusion time and the area, hence the curve crosses the origin (Fig.1 c). The presence of impermeable obstacles constrains the diffusion and increases the apparent time to cross a given observation area, thus the slope of the FCS diffusion law is higher, but the origin (0,0) is still crossed. Notably, the presence of confinement affecting the lateral diffusion is revealed by a deviation of the intercept on the time axis $t_0$ from the origin (Fig.1 c). Extrapolating the experimental curve to the intercept with the time axis, hindered diffusion due to nanodomains is regarded as a positive intercept on the time axis, while the meshwork model is related to a negative intercept. This approach was established on diffraction-limited confocal microscopes, where the FCS diffusion law for the transferrin receptor TfR-GFP (known to interact with the cytoskeleton meshwork) yielded a negative $t_0$ value, while that of the fluorescent ganglioside GM1 exhibited a positive $t_0$ value.[20,51] By extrapolating to the origin, the FCS diffusion laws can predict the occurrence of membrane heterogeneities affecting the lateral diffusion at spatial scales well beyond the optical resolution. However, the size of lipid rafts is expected to be around 10-100 nm[18,19,52], so their areas are 5 to 500 x smaller than the smallest diffraction-limited observation area on confocal microscopes. Reducing this gap between optical resolution and the size of lipid rafts to gain better insights on membrane organization at the nanoscale is currently a field of active research.

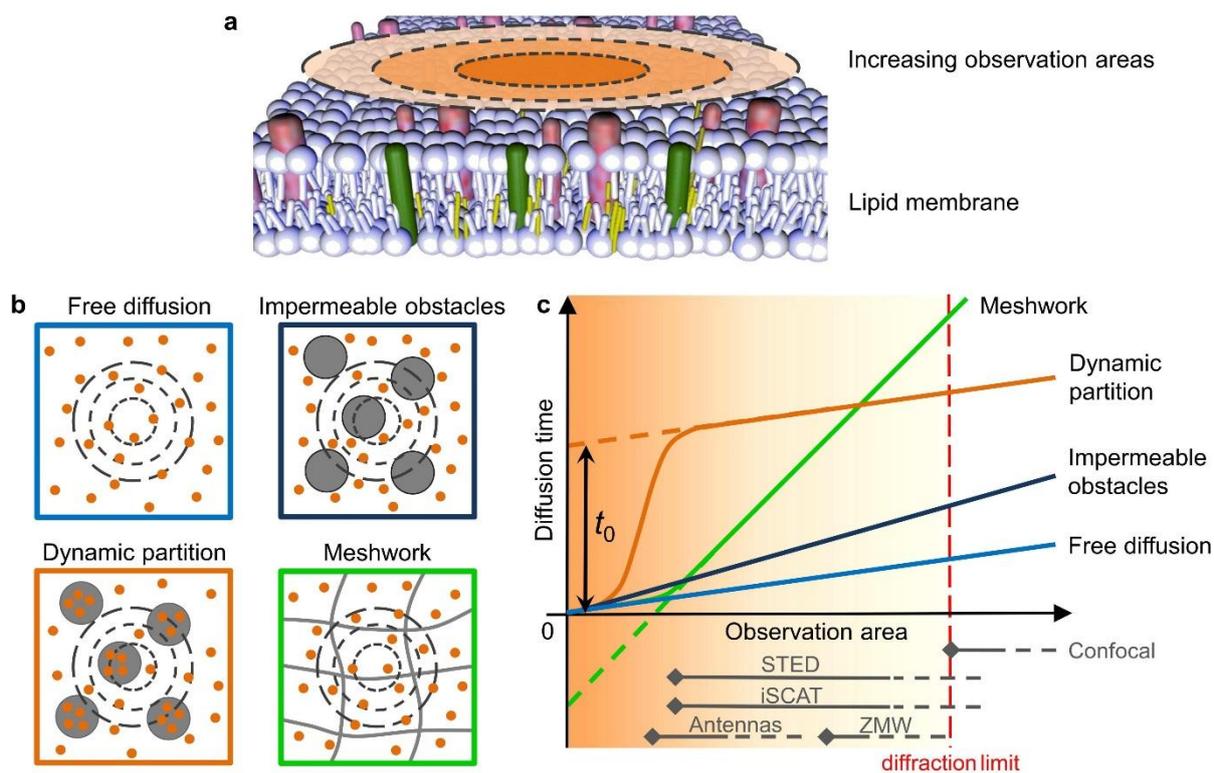

**Figure 1. Principles of FCS diffusion laws to reveal biomembrane organization at the nanoscale.** (a) FCS diffusion laws are constructed by measuring the diffusion times of molecules traversing illumination areas of increasing sizes. (b) Different diffusion models depending on the membrane organization can be distinguished by varying the illumination areas. Molecules can freely diffuse on the membrane or show hindered diffusion due to their dynamic partitioning into nanodomains or due to the cortical actin meshwork. (c) FCS diffusion laws showing diffusion times versus observation area. The type of diffusion is retrieved by extrapolation of the curves through the y-axis intercept $t_0$. Free diffusion and impermeable obstacles are characterized by $t_0 = 0$, while a positive $t_0$ intercept indicates the presence of nanodomains transiently trapping the molecular probe. A negative $t_0$ intercept relates to a meshwork of barriers separating adjacent domains. The observation areas accessible with various super-resolution techniques are indicated as grey lines.

With the advent of super-resolution optical microscopy approaches such as single molecule localization methods[53–55], stimulated emission depletion (STED) microscopy[56–58] and near-field scanning optical microscopy (NSOM)[8,59–62], it is now becoming clearer that lipids and proteins can indeed organize in nanometric compartments on the cell membrane, albeit a consensus in terms of their sizes and dynamics has not yet been reached. In terms of dynamic measurements at the nanoscale, NSOM has been combined with FCS to show anomalous diffusion of ganglioside GM1 on living cell membranes at sizes smaller than 120nm.[63] STED has been also combined with FCS to explore the nanoscale dynamics occurring in lipid membranes.[22,64–68] Notably, STED-FCS experiments on living cell membranes revealed that unlike phosphoglycerolipids, sphingolipids and GPI-APs are transiently trapped in cholesterol-mediated molecular complexes of sub-20 nm dimensions.[22] These landmark results rely on the key benefits of STED-FCS to provide sub-diffraction spatial (super)resolution, tunable observation areas, and microsecond temporal resolution. STED-FCS was also applied to study ternary lipid-cholesterol model membranes featuring microscopic liquid-liquid phase separation into Ld and Lo phases, without observing any direct evidence of the presence of nanoscopic domains at the spatial scales down to 40nm.[65] However, nanoscale assemblies smaller than the 40 nm minimum STED resolution may still exist without being observed. Further recent technical developments extended the FCS functionalities to scanning STED-FCS[67] being specially adapted to membrane studies of slow diffusion, and to fluorescence lifetime filtering STED-FLCS.[68] Recently, Basu and coworkers detected dynamic heterogeneities at length scales of ~ 100-150 nm in binary phospholipid-cholesterol bilayers of high cholesterol content (50 %) by applying STED-FCS (with a resolution of ~ 80nm).[69] The occurrence of these heterogeneities in binary model membranes showing no macroscopic phase separation indicates that the domain formation is driven by cholesterol packing and influenced by the phospholipid type. However, the high cholesterol content (50 %) used in the binary mixtures complicates a direct comparison to cellular membranes.

Advanced SPT[3,26,70] and the recently introduced high-speed SPT interferometric scattering microscopy (iSCAT) technique[71] enable direct visualization of single particle trajectories. iSCAT microscopy allows nowadays nanometer localization precision together with microsecond time resolution by means of using 20-40nm gold nanoparticles as labeling probes.[72,73] Recent high-speed SPT experiments using 20 nm gold beads attached to individual lipids in multicomponent model membranes showed anomalous diffusion in the Lo phase consistent with the occurrence of nanoscale heterogeneities, while homogeneous lipid diffusion was observed in the Ld phase.[74] The estimated sizes of the nanodomains in the Lo phase varied between 20 to 40 nm with lipid trapping times inside the domains below 1 ms. iSCAT thus constitutes an attractive tool to investigate dynamic biophysical processes in mimetic systems at the nanoscale, yet additional investigations are still required to rule out potential artifacts related to the large size of the gold nanoparticle label with respect to the lipids under study.

Beside these enormous progresses in super-resolution microscopy and single-molecule dynamic approaches, advances from the nanophotonics field have led to the concept of photonic nanostructures to confine light on a subwavelength scale and reach sub-diffraction observation areas in FCS.[75–77] A conceptually simple yet powerful approach uses single nanometric apertures milled in a metallic film also known as zero-mode waveguides (ZMW) to confine the illumination spot directly in the sample plane.[78] Typically, the apertures have radii between 50 to 250 nm and are milled in an opaque aluminum film covering a glass coverslip.[79,80] Their combination with FCS has been used to probe model lipid membranes[81,82] and living cell membranes[83–86], revealing for instance that fluorescent chimeric ganglioside proteins partition into 30 nm structures within the cell membrane.[86] While the ZMW approach is very efficient at confining light within nanospots of diameters between 100 to 200 nm, this technique has difficulties reaching spot sizes below 80 nm. Indeed, the FCS signal-to-noise ratio rapidly deteriorates for ZMW diameters below 100 nm as a consequence of fluorescence quenching induced by the metallic aperture edges.[87] An additional issue affecting the use of ZMWs for living cell membrane studies is the lack of control on the membrane invagination into the aperture. This problem has been addressed by introducing a planarization procedure in the nanofabrication process filling the aperture volume with fused silica.[88,89] Thanks to the absence of a height difference between the ZMW and the surrounding metal layer, the cells can lie on a nearly perfectly flat surface. The best results achieved so far reach a nanospot diameter of 60 nm and microsecond resolution.[89]

The concept of resonant optical nanoantennas has been introduced to further confine the excitation light down to sub-20 nm scales.[90–92] Optical nanoantennas are metallic (plasmonic) nanostructures that localize and enhance the incident optical radiation into highly confined nanometric regions (plasmonic hotspots), leading to greatly enhanced light-matter interactions.[93,94] Over thousand-fold enhancement of the single molecule fluorescence signal was reported with lithographically fabricated gold nanoantennas in the shape of bowties[95], with dimers of gold nanoparticles assembled with DNA origami[96,97] and at the apex of single gold nanorods.[98,99]. Recent advances in nanofabrication using colloid nanosphere lithography combined with plasma processing[100] and nanostencil lithography[101] enable nowadays large scale production of reproducible nanoantennas with narrow gaps as required for the study of the plasma membrane of living cells. However, the applications of plasmonic antennas to living cells remain scarce. Plasmon-enhanced fluorescence was recently observed inside living bacterial cell membranes[102], highlighting the need to develop well-tuned substrates to maximize fluorescence enhancement and signal-to-noise ratio. In the highly active field of biosensing in the context of nanomedicine[103] plasmonic antennas have enabled to perform Raman spectroscopy in a microfluidic device on the single cell level[104] or to detect single amino acid mutations in breast cancer cells.[105]

A major issue limiting the use of optical nanoantennas for living cell membrane studies is the efficient rejection of the background fluorescence light originating from the molecules that are sufficiently away (tens of nanometers) from the antenna hotspot but still within the diffraction limited confocal volume.[106] For antennas made of individual nanoparticles[107–111] or dimers of nanoparticles[112–114] deposited on a glass substrate, the fluorescence background can be significantly larger than the antenna-enhanced fluorescence signal from the plasmonic hotspot, challenging single molecule detection and FCS using nanoantennas. The initial approach to deal with this challenge employed low quantum yield emitters (quantum yield below 8 %) leading to maximizing the apparent fluorescence enhancement while minimizing the background[110–112,114] Another solution relies on time gating and lifetime filtering, taking advantage of the reduced lifetime of the emitters in the vicinity of the plasmonic hotspot.[115] A third approach, and the one that we will detail further in this Perspective, uses a dedicated antenna design termed "antenna-in-box".[116,117]

The "antenna-in-box" platform features a metal dimer nanogap antenna centered inside a nanoaperture and is specifically designed for FCS and single molecule analysis at physiologically relevant (micromolar) concentrations (Fig. 2). The central nanogap antenna provides the nanoscale plasmonic hotspot, while the surrounding metal cladding screens the fluorescence background by preventing the excitation of the molecules diffusing away from the nanogap.[116] A challenge associated with classical nanofabrication techniques such as focused ion beam milling or electron beam lithography is that the region of maximum field localization is buried into the nanostructure and not directly accessible for fluorescent emitters embedded in a membrane. We recently overcame this issue by combining electron beam lithography with planarization, etch back and template stripping.[118] The planarization strategy fills the aperture volume with a transparent polymer, yielding a flat top surface (of a planarity better than 3 nm, see Fig. 2 c), compatible with membrane studies on living cells. Possible curvature induced effects on the cell membrane are thus avoided.[88,89] The etch back approach produces reproducible arrays of nanoantennas with controlled gap sizes and sharp edges. With a gap size of 10 nm, the antenna gap area can be as small as 300 nm$^2$ (Fig. 2 d,e), realizing a reduction of 200 x as compared to the diffraction-limited confocal area. Lastly, the template stripping flips the plasmonic hotspot to the top surface and place it in the immediate vicinity of the cell membrane. Owing to these nanofabrication advances, planar plasmonic nanoantennas drastically improve optical performance leading to fluorescence enhancement factors above 10,000 x (for crystal violet dyes of 2 % quantum yield) and detection volumes in the zeptoliter range.[118]

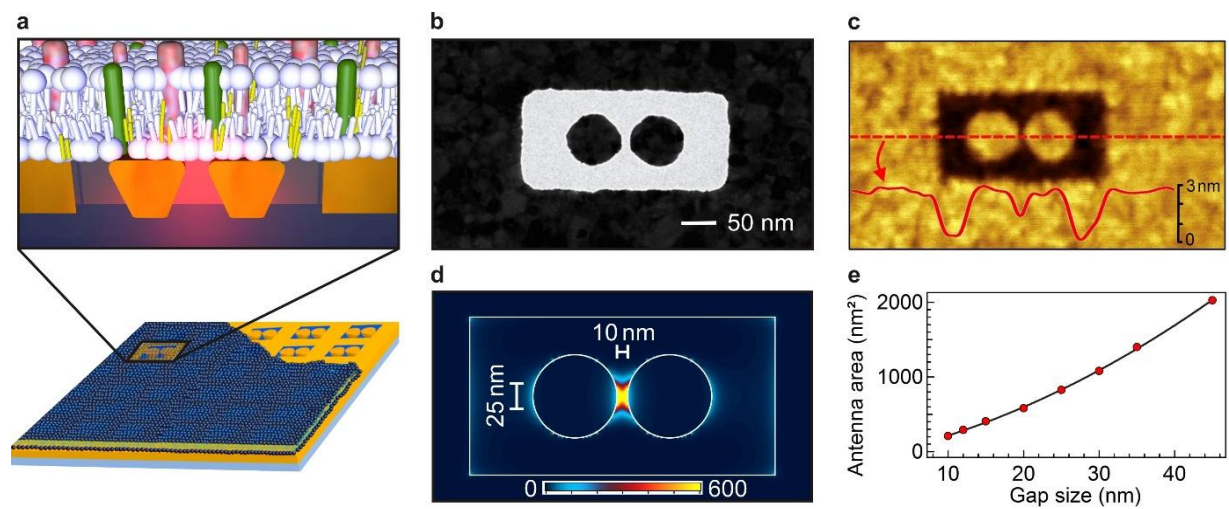

**Figure 2. Planar optical nanoantennas to investigate lipid biological membranes.** (a) Schematics of the experimental arrangement. Arrays of planar optical antennas are fabricated by electron beam lithography. Each antenna consists of a dimer of gold nanoparticles separated by a nanometric gap embedded in a polymer filling a rectangular aperture. The gold dimer confines the excitation light into a nanometric hotspot while the metal cladding prevents direct excitation of the surrounding fluorescently labeled membrane. The lipid membrane is directly prepared on top of the planar nanoantennas. (b) Transmission electron microscope (TEM) image of a representative antenna with 10 nm gap size. (c) Atomic force microscope (AFM) image of the antenna sample top surface. The topography profile (red curve) along the antenna axis shows variations in height below 3 nm across the antenna. (d) Finite-difference time-domain (FDTD) simulations of the electric field intensity profile within a 10 nm gap antenna for an illumination wavelength of 633 nm. The color scale indicates the enhancement of the local excitation intensity. (e) Simulations of the observation areas as a function of the gap size. Each observation area is computed as the product of the gap size times the full width at half maximum of the intensity profile along the direction perpendicular to the antenna main axis.

We have recently used these planar plasmonic nanoantennas in combination with FCS to assess the dynamic nanoscale organization of mimetic biological membranes.[119] As already mentioned, tertiary model lipid membranes composed of phospholipids, sphingolipids and cholesterol separate into stable coexisting Lo and Ld phases which are microscopic in size and easily observable by diffraction-limited optical microscopy.[32–35,120,121] However, there are intriguing indications that the microscopically homogeneous Lo and Ld phases in model lipid membranes might in fact be also heterogeneously organized at the nanometer scale, resembling the scenario occurring in living cells.[18,74,122] Indeed, atomistic and coarse-grained simulations have predicted the existence of highly transient lipid clusters around 10 nm in size and with microsecond lifetimes within both phases, namely raft and non-raft domains of multicomponent membranes.[123] In support of these simulations, recent NMR experiments demonstrated that a significant amount of saturated lipids and cholesterol is present in the Ld phase, and that unsaturated lipids are also found in the Lo phase, strongly pointing towards the existence of nanoscopic assemblies in both phases.[122] However, other workers have observed the presence of transient nanoscopic domains only in the Lo phase[74,124,125], while others have shown the occurrence of nanoscopic heterogeneities in the Ld phase.[126,127] Finally, a STED-FCS study showed no nanodomain formation down to 40nm (set by the STED resolution) in any of the phases, indicating that both phases are homogeneously distributed.[65]

To investigate the potential existence of transient nanoscopic heterogeneities within the microscopically homogeneous Lo and Ld phases in model lipid membranes we thus took advantage of the planar plasmonic nanogap antenna platform combined with FCS at various nanoscale illumination areas.[119] We analyzed the diffusion of individual DiD fluorescent molecules inserted in lipid bilayers composed of the unsaturated phospholipid 1,2-dioleoyl-sn-glycero-3-phosphocholine (DOPC) alone, DOPC in combination with sphingomyelin (SM) (1:1 molar proportions) and of the two ternary mixtures of DOPC, SM (1:1) with addition of 10 or 20 mol % cholesterol (Chol) (Fig. 3a). Using nanoantennas of gap sizes from 10 to 45 nm, the FCS diffusion laws could be extended down to areas of a few hundreds of $nm^2$ (Fig. 3b). The results for the diffusion of the dye in pure DOPC membranes indicated free Brownian diffusion down to the nanoscale, as expected for a homogeneous lipid distribution. Importantly, these results confirm that the planar nanoantenna platform does not introduce any artifacts hindering the diffusion of fluorescent dyes. In the presence of cholesterol, microscopic phase separation occurred into the Lo and Ld phases, enabling the investigation of putative nanoscopic heterogeneities in both phases. Interestingly, the FCS diffusion laws for both the Lo and Ld phase displayed positive y-axis intercepts $t_0$ significantly deviating from free Brownian diffusion (Fig. 3b). These results indicate the presence of transient nanoscopic domains in both the Ld and Lo phases of ternary lipid mixtures with sizes about 10 nm and short characteristic times around 30 μs for the Ld, and 100 μs for the Lo phases (Fig. 3c).[119] The extremely short-lived average residence time for the heterogeneities in the Ld phase is most likely the reason why they have not been detected before even with high-resolution SPT.[74] Although the plasma membrane of living cells bears a much higher complexity, these nanoscale assemblies in lipid model membranes might illustrate a general underlying principle setting the basis for lipid raft formation in living cells.

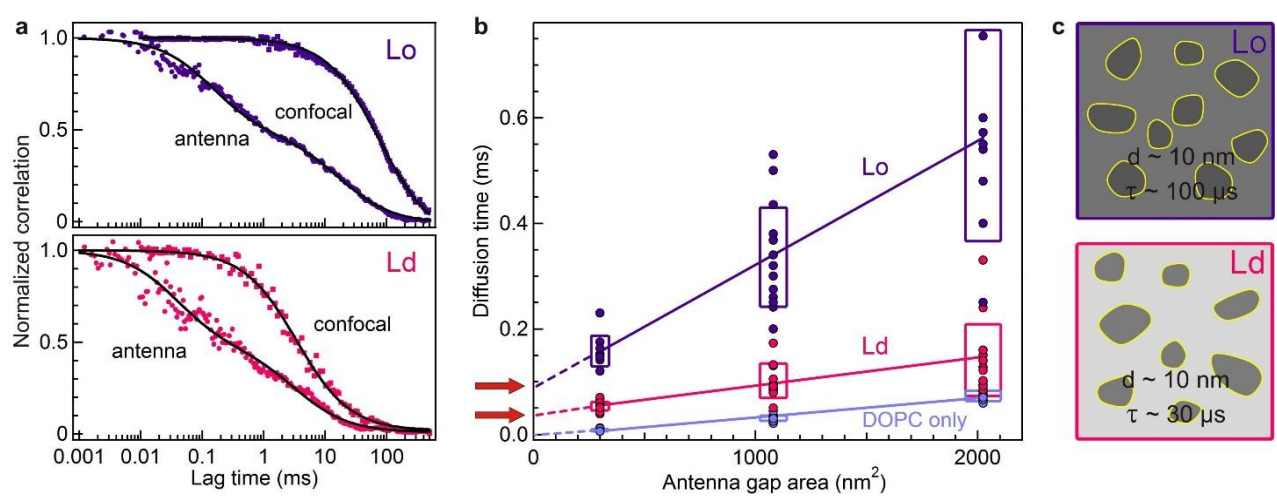

**Figure 3. Transient nanodomains in biological model membranes of ternary lipid mixtures resolved by optical nanoantennas.** (a) FCS correlation curves for DOPC:SM(1:1) + Cholesterol 20 mol % lipid mixtures recorded for the liquid ordered (Lo) and liquid disordered (Ld) phases in the confocal setup and with a 12 nm gap antenna. (b) FCS diffusion laws at the nanoscale for the ternary lipid mixture DOPC:SM(1:1) + Cholesterol 20 mol % in the Lo and Ld phases. The results for pure DOPC bilayers are shown for comparison. (c) The positive y-axis intercept $t_0$ (highlighted by red arrows) in (b) reveals the existence of transient nanoscopic domains in both the Lo and Ld phases for lipid mixtures containing cholesterol.

To extend the applications of planar nanoantennas to membrane studies, we recently used these platforms in combination with FCS to measure the nanoscale dynamics of different lipids in fully intact living cells.[128] For this, living Chinese hamster ovary (CHO) cells were incubated in a cell culture well on the antenna platform at 37 °C so that the cells could nicely adhere on the nanoantenna substrate (Fig. 4 a).[128] We discuss here the results obtained by measuring the sphingomyelin (SM) lipid analog labeled with the fluorescent dye Atto647N before and after methyl-β-cyclodextrin (MCD) treatment. MCD depletes cholesterol from the cell membrane, which is expected to play a significant role in the formation of lipid rafts. With the diffraction-limited resolution of the confocal microscope, the typical fluorescence bursts (Fig.4 b) and FCS traces (Fig.4 c) for both SM before and after MCD treatment (MCD-SM) showed barely distinguishable features. In stark contrast, clear differences between the SM and MCD-SM signals were observed with a 12 nm hotspot from the nanoantenna, indicating the influence of cholesterol in hindering the diffusion of SM.

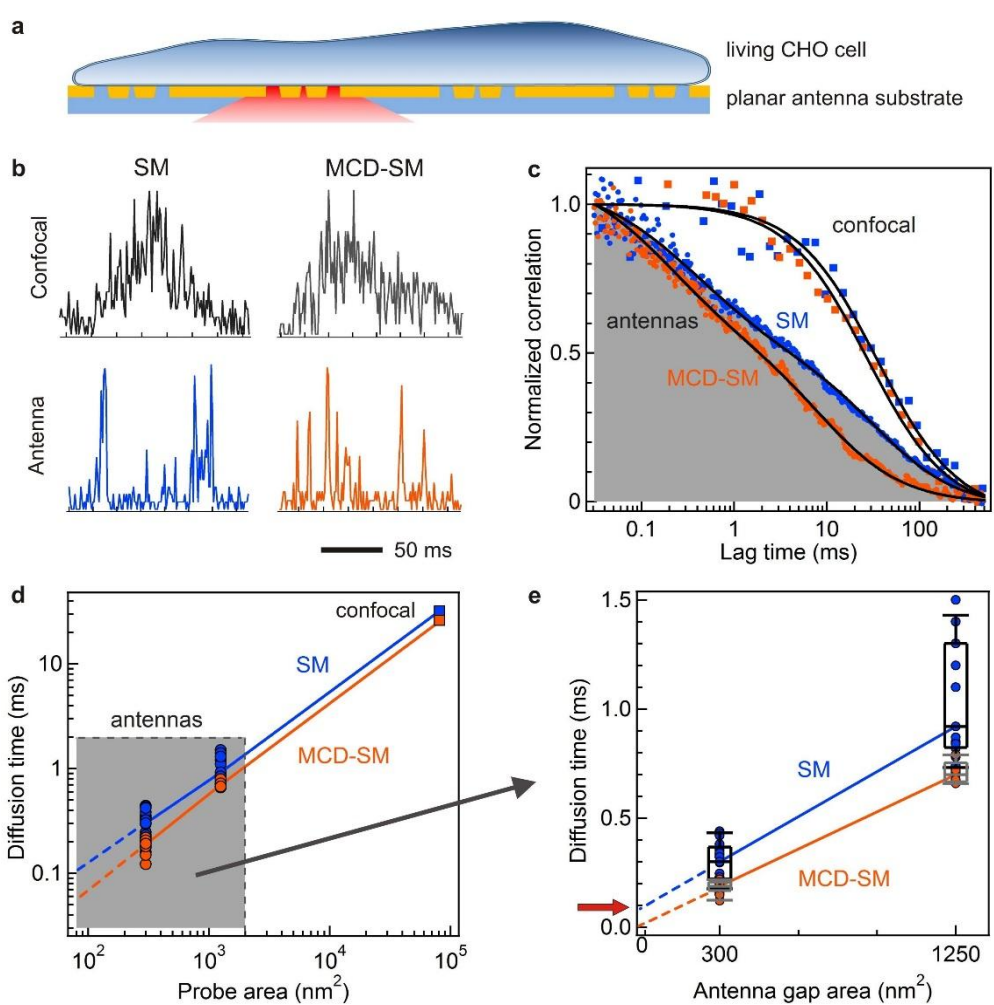

**Figure 4. Lipid membrane organization in living cells probed by optical nanoantennas.** (a) Experimental scheme: live CHO cells are seeded and grown directly on the planar antenna substrate. (b) Example of fluorescence bursts and FCS correlation traces (c) for sphingomyelin (SM) before and after cholesterol depletion using MCD treatment, recorded in the confocal setup and with a 12 nm gap antenna. (d,e) FCS diffusion laws for SM before and after MCD treatment. Linear fits through the median values (continuous lines) are extrapolated through the y-axis intercept $t_0$ (dashed lines, red arrow). The box plots in (e) represent the 25th, 50th and 75th percentiles while the bars indicate the 10th and 90th percentiles (for 48 different antennas).

We further plotted the FCS diffusion laws recorded on nanoantennas for SM and SM after MCD treatment (Fig. 4 d,e). The slope of the fitted curves allows to determine the diffusion coefficients to $D_{SM} = 0.38 \pm 0.19$ µm$^2$/s and $D_{MCD-SM} = 0.46 \pm 0.07$ µm$^2$/s, which are consistent with the confocal measurements. Extrapolating the fits to estimate the time-axis intercepts, we found for SM a positive $t_{0,SM} \sim 110 \pm 80$ µs, while after MCD treatment, the intercept comes close to zero with $t_{0, MCD-SM} \sim 20 \pm 15$ µs. It should be pointed out that a positive time-axis intercept $t_0$ does *not* exactly correspond to the trapping time of the nanodomains. As shown by Ruprecht and coworkers[50], in the case of immobile nanodomains and an exponential distribution of trapping and diffusion times, the time offset $t_0$ is the product of the trapping time $\tau_{trap}$ times the fraction β of trapped fluorophores: $t_0 = \beta \tau_{trap}$. Likewise, the effective diffusion coefficient $D_{eff}$ measured from the FCS diffusion laws can be expressed $D_{eff} = (1-\beta) D_{free}$ where $D_{free}$ is the diffusion coefficient for the free dye. Using the experimental values measured for SM before and after MCD treatment (and substituting in the previous equations $D_{SM} = D_{eff}$ and $D_{MCD-SM} = D_{free}$), we obtain β=0.17 and $\tau_{trap}$ =0.6 ms. A slightly modified set of equations allows to take into account also the mobility of the nanodomains.[50] Assuming that the diffusion coefficient for the nanodomains is ten times slower than for free diffusion $D_{trap} = D_{free} /10$, we obtain slightly modified values for the trapped fraction and trapping time, i.e., β =0.19 and $\tau_{trap.}$ =0.9 ms. These results stand in good agreement with the 1-2~ms trapping time inferred from STED-FCS using an anomalous diffusion fitting (Fig. S3 of Ref. [22]). Altogether, these results indicate the occurrence of cholesterol-dependent nanodomains hindering SM diffusion in living cell membranes with sub-millisecond characteristic times and typical sizes below 10 nm (as inferred from the smallest gap size of our antennas). These experimental observations extend the previous works[20,51] on FCS diffusion laws to the nanoscale dimension far below the diffraction limit. Taking advantage of the narrow gap sizes down to 10 nm, this approach also allows to explore membrane organization on areas below the $10^{-3}$-$10^{-4}$ nm$^2$ spatial scale probed by STED-FCS.[22] The nanodomain characteristics stand in good agreement with the predictions from stochastic models[18,123,129], and with the current understanding of lipid rafts as highly transient and fluctuating nanoscale assemblies of sterol and sphingolipids.[2,7,52]

In conclusion, planar plasmonic nanoantennas with accessible surface nanogaps offer a promising new approach to investigate the dynamic nanoscale organization of living cell membranes. The proof-of-principle demonstrations on model lipid membranes[119] and CHO cell membranes[128] constitute a significant step forward in our ability to address native biological membranes with ultrahigh spatiotemporal resolution at the nanometer and microsecond scales. The nanoantennas provide an encouraging outlook to investigate the dynamics and interactions of lipids and raft-associated proteins and their recruitment into molecular complexes. These studies will ultimately improve our understanding of the cell membrane organization and its link to the cell's function.

Working on multicomponent mimetic biological membranes permits to investigate the nanoscale dynamic organization of biological membranes and its impact in biological function in a controllable manner. Future directions involve the addition of more complex components into the mimetic system such as membrane signaling proteins, components of the glycan network or a cortical actin mesh. The envisioned next steps in living cells will explore the native influence of the adjacent inner and outer environment (the cortical actin cytoskeleton and the glycan network, respectively) on templating the dynamic nanoscale organization of the plasma membrane. Reaching these goals will also require pushing the nanoantenna technology even further, to narrow the antenna gap, sharpen the metal edges, improve the overall reproducibility over the full antenna arrays and enabling multiplexed, parallel detection from hundreds of antennas simultaneously. Additional challenges comprise the development of antennas with broadband resonance enabling multi-color fluorescence detection. Altogether, this outlook preludes a new class of biomolecular studies with ultrahigh spatial and temporal resolutions, reaching the long-awaited goal of nanometer spatial precision combined with microsecond temporal resolution and with full biocompatibility.

**Additional Information**

The authors declare no competing financial interests.

**Acknowledgements**

The research leading to these results has received funding from the European Commission's Seventh Framework Programme (FP7-ICT-2011-7) under grant agreements ERC StG 278242 (ExtendFRET), 288263 (NanoVista), Spanish Ministry of Economy and Competitiveness ("Severo Ochoa" Programme for Centres of Excellence in R&D (SEV-2015-0522) and FIS2014-56107-R), Fundació CELLEX (Barcelona) and CERCA Programme/Generalitat de Catalunya. P.M.W is supported by the ICFOstepstone Fellowship, a COFUND Doctoral Programme of the Marie-Sklodowska-Curie-Action of the European Commission. R.R. is supported by the Erasmus Mundus Doctorate Program Europhotonics (Grant 159224-1-2009-1-FR-ERA MUNDUS-EMJD).

**References**

(1) Brown, D. A.; London, E. Functions of Lipid Rafts in Biological Membranes. *Annu. Rev. Cell Dev. Biol.* **1998**, *14*, 111–136.

(2) Lingwood, D.; Simons, K. Lipid Rafts as a Membrane-Organizing Principle. *Science* **2010**, *327*, 46–50.

(3) Kusumi, A.; Nakada, C.; Ritchie, K.; Murase, K.; Suzuki, K.; Murakoshi, H.; Kasai, R. S.; Kondo, J.; Fujiwara, T. Paradigm Shift of the Plasma Membrane Concept from the Two-Dimensional Continuum Fluid to the Partitioned Fluid: High-Speed Single-Molecule Tracking of Membrane Molecules. *Annu. Rev. Biophys. Biomol. Struct.* **2005**, *34*, 351–378.

(4) Gowrishankar, K.; Ghosh, S.; Saha, S.; C., R.; Mayor, S.; Rao, M. Active Remodeling of Cortical Actin Regulates Spatiotemporal Organization of Cell Surface Molecules. *Cell* **2012**, *149*, 1353–1367.

(5) Sezgin, E.; Levental, I.; Mayor, S.; Eggeling, C. The Mystery of Membrane Organization: Composition, Regulation and Roles of Lipid Rafts. *Nat. Rev. Mol. Cell Biol.* **2017**, *18*, 361–374.

(6) Simons, K.; Ikonen, E. Functional Rafts in Cell Membranes. *Nature* **1997**, *387*, 569.

(7) Mayor, S.; Rao, M. Rafts: Scale-Dependent, Active Lipid Organization at the Cell Surface: Raft Hypothesis. *Traffic* **2004**, *5*, 231–240.

(8) van Zanten, T. S.; Cambi, A.; Koopman, M.; Joosten, B.; Figdor, C. G.; Garcia-Parajo, M. F. Hotspots of GPI-Anchored Proteins and Integrin Nanoclusters Function as Nucleation Sites for Cell Adhesion. *Proc. Natl. Acad. Sci.* **2009**, *106*, 18557–18562.

(9) Raghu, H.; Sodadasu, P. K.; Malla, R. R.; Gondi, C. S.; Estes, N.; Rao, J. S. Localization of uPAR and MMP-9 in Lipid Rafts Is Critical for Migration, Invasion and Angiogenesis in Human Breast Cancer Cells. *BMC Cancer* **2010**, *10*, 647.

(10) Lingwood, D.; Binnington, B.; Róg, T.; Vattulainen, I.; Grzybek, M.; Coskun, Ü.; Lingwood, C. A.; Simons, K. Cholesterol Modulates Glycolipid Conformation and Receptor Activity. **2011**, *7*, 260.

(11) Rios, F. J. O.; Ferracini, M.; Pecenin, M.; Koga, M. M.; Wang, Y.; Ketelhuth, D. F. J.; Jancar, S. Uptake of oxLDL and IL-10 Production by Macrophages Requires PAFR and CD36 Recruitment into the Same Lipid Rafts. *PLOS ONE* **2013**, *8*, e76893.

(12) Laganowsky, A.; Reading, E.; Allison, T. M.; Ulmschneider, M. B.; Degiacomi, M. T.; Baldwin, A. J.; Robinson, C. V. Membrane Proteins Bind Lipids Selectively to Modulate Their Structure and Function. *Nature* **2014**, *510*, 172–175.

(13) Farnoud, A. M.; Toledo, A. M.; Konopka, J. B.; Del Poeta, M.; London, E. Raft-Like Membrane Domains in Pathogenic Microorganisms. *Curr. Top. Membr.* **2015**, *75*, 233–268.

(14) Larsen, J. B.; Jensen, M. B.; Bhatia, V. K.; Pedersen, S. L.; Bjørnholm, T.; Iversen, L.; Uline, M.; Szleifer, I.; Jensen, K. J.; Hatzakis, N. S.; *et al.* Membrane Curvature Enables N-Ras Lipid Anchor Sorting to Liquid-Ordered Membrane Phases. **2015**, *11*, 192.

(15) Varshney, P.; Yadav, V.; Saini, N. Lipid Rafts in Immune Signalling: Current Progress and Future Perspective. *Immunology* **2016**, *149*, 13–24.

(16) Munro, S. Lipid Rafts: Elusive or Illusive? *Cell* **2003**, *115*, 377–388.

(17) Pike, L. J. Rafts Defined: A Report on the Keystone Symposium on Lipid Rafts and Cell Function. *J. Lipid Res.* **2006**, *47*, 1597–1598.

(18) Hancock, J. F. Lipid Rafts: Contentious Only from Simplistic Standpoints. *Nat Rev Mol Cell Biol* **2006**, *7*, 456–462.

(19) Marguet, D.; Lenne, P.; Rigneault, H.; He, H. Dynamics in the Plasma Membrane: How to Combine Fluidity and Order. *EMBO J.* **2006**, *25*, 3446.

(20) Lenne, P.-F.; Wawrezinieck, L.; Conchonaud, F.; Wurtz, O.; Boned, A.; Guo, X.-J.; Rigneault, H.; He, H.-T.; Marguet, D. Dynamic Molecular Confinement in the Plasma Membrane by Microdomains and the Cytoskeleton Meshwork. *EMBO J.* **2006**, *25*, 3245–3256.

(21) Jacobson, K.; Mouritsen, O. G.; Anderson, R. G. W. Lipid Rafts: At a Crossroad between Cell Biology and Physics. *Nat Cell Biol* **2007**, *9*, 7–14.

(22) Eggeling, C.; Ringemann, C.; Medda, R.; Schwarzmann, G.; Sandhoff, K.; Polyakova, S.; Belov, V. N.; Hein, B.; von Middendorff, C.; Schönle, A.; *et al.* Direct Observation of the Nanoscale Dynamics of Membrane Lipids in a Living Cell. *Nature* **2009**, *457*, 1159–1162.

(23) Klotzsch, E.; Schütz, G. J. A Critical Survey of Methods to Detect Plasma Membrane Rafts. *Philos. Trans. R. Soc. B Biol. Sci.* **2013**, *368*, 20120033.

(24) Singer, S. J.; Nicolson, G. L. The Fluid Mosaic Model of the Structure of Cell Membranes. *Science* **1972**, *175*, 720.

(25) Goswami, D.; Gowrishankar, K.; Bilgrami, S.; Ghosh, S.; Raghupathy, R.; Chadda, R.; Vishwakarma, R.; Rao, M.; Mayor, S. Nanoclusters of GPI-Anchored Proteins Are Formed by Cortical Actin-Driven Activity. *Cell* **2008**, *135*, 1085–1097.

(26) Fujiwara, T.; Ritchie, K.; Murakoshi, H.; Jacobson, K.; Kusumi, A. Phospholipids Undergo Hop Diffusion in Compartmentalized Cell Membrane. *J. Cell Biol.* **2002**, *157*, 1071–1082.

(27) Lajoie, P.; Goetz, J. G.; Dennis, J. W.; Nabi, I. R. Lattices, Rafts, and Scaffolds: Domain Regulation of Receptor Signaling at the Plasma Membrane. *J. Cell Biol.* **2009**, *185*, 381–385.

(28) Subramaniam, A. B.; Guidotti, G.; Manoharan, V. N.; Stone, H. A. Glycans Pattern the Phase Behaviour of Lipid Membranes. *Nat. Mater.* **2012**, *12*, 128–133.

(29) Groves, J. T. Cell Membranes: Glycans' Imprints. *Nat. Mater.* **2013**, *12*, 96–97.

(30) Garcia-Parajo, M. F.; Cambi, A.; Torreno-Pina, J. A.; Thompson, N.; Jacobson, K. Nanoclustering as a Dominant Feature of Plasma Membrane Organization. *J. Cell Sci.* **2014**, *127*, 4995–5005.

(31) Blouin, C. M.; Hamon, Y.; Gonnord, P.; Boularan, C.; Kagan, J.; Viaris de Lesegno, C.; Ruez, R.; Mailfert, S.; Bertaux, N.; Loew, D.; *et al.* Glycosylation-Dependent IFN-γR Partitioning in Lipid and Actin Nanodomains Is Critical for JAK Activation. *Cell 166*, 920–934.

(32) Dietrich, C.; Bagatolli, L. A.; Volovyk, Z. N.; Thompson, N. L.; Levi, M.; Jacobson, K.; Gratton, E. Lipid Rafts Reconstituted in Model Membranes. *Biophys. J.* **2001**, *80*, 1417–1428.

(33) Veatch, S. L.; Keller, S. L. Organization in Lipid Membranes Containing Cholesterol. *Phys. Rev. Lett.* **2002**, *89*, 268101.

(34) Kahya, N.; Scherfeld, D.; Bacia, K.; Poolman, B.; Schwille, P. Probing Lipid Mobility of Raft-Exhibiting Model Membranes by Fluorescence Correlation Spectroscopy. *J. Biol. Chem.* **2003**, *278*, 28109–28115.

(35) Chiantia, S.; Ries, J.; Kahya, N.; Schwille, P. Combined AFM and Two-Focus SFCS Study of Raft-Exhibiting Model Membranes. *ChemPhysChem* **2006**, *7*, 2409–2418.

(36) Simons, K.; Vaz, W. L. C. Model Systems, Lipid Rafts, and Cell Membranes. *Annu. Rev. Biophys. Biomol. Struct.* **2004**, *33*, 269–295.

(37) Tamm, L. K.; McConnell, H. M. Supported Phospholipid Bilayers. *Biophys. J.* **1985**, *47*, 105–113.

(38) Sezgin, E.; Kaiser, H.-J.; Baumgart, T.; Schwille, P.; Simons, K.; Levental, I. Elucidating Membrane Structure and Protein Behavior Using Giant Plasma Membrane Vesicles. *Nat Protoc.* **2012**, *7*, 1042–1051.

(39) Baumgart, T.; Hammond, A. T.; Sengupta, P.; Hess, S. T.; Holowka, D. A.; Baird, B. A.; Webb, W. W. Large-Scale Fluid/Fluid Phase Separation of Proteins and Lipids in Giant Plasma Membrane Vesicles. *Proc. Natl. Acad. Sci.* **2007**, *104*, 3165–3170.

(40) Lingwood, D.; Ries, J.; Schwille, P.; Simons, K. Plasma Membranes Are Poised for Activation of Raft Phase Coalescence at Physiological Temperature. *Proc. Natl. Acad. Sci.* **2008**, *105*, 10005–10010.

(41) Hammond, A. T.; Heberle, F. A.; Baumgart, T.; Holowka, D.; Baird, B.; Feigenson, G. W. Crosslinking a Lipid Raft Component Triggers Liquid Ordered-Liquid Disordered Phase Separation in Model Plasma Membranes. *Proc. Natl. Acad. Sci. U. S. A.* **2005**, *102*, 6320–6325.

(42) Kenworthy, A. K. Fluorescence Recovery After Photobleaching Studies of Lipid Rafts. In *Lipid Rafts*; McIntosh, T. J., Ed.; Humana Press: Totowa, NJ, 2007; pp. 179–192.

(43) Chen, Y.; Lagerholm, B. C.; Yang, B.; Jacobson, K. Methods to Measure the Lateral Diffusion of Membrane Lipids and Proteins. *Anal. Methods Sci. Lipidomics Membr. Organ. Protein-Lipid Interact.* **2006**, *39*, 147–153.

(44) Dietrich, C.; Yang, B.; Fujiwara, T.; Kusumi, A.; Jacobson, K. Relationship of Lipid Rafts to Transient Confinement Zones Detected by Single Particle Tracking. *Biophys. J. 82*, 274–284.

(45) Bacia, K.; Kim, S. A.; Schwille, P. Fluorescence Cross-Correlation Spectroscopy in Living Cells. *Nat Meth* **2006**, *3*, 83–89.

(46) He, H.-T.; Marguet, D. Detecting Nanodomains in Living Cell Membrane by Fluorescence Correlation Spectroscopy. *Annu. Rev. Phys. Chem.* **2011**, *62*, 417–436.

(47) Maiti, S.; Haupts, U.; Webb, W. W. Fluorescence Correlation Spectroscopy: Diagnostics for Sparse Molecules. *Proc. Natl. Acad. Sci.* **1997**, *94*, 11753–11757.

(48) Yechiel, E.; Edidin, M. Micrometer-Scale Domains in Fibroblast Plasma Membranes. *J. Cell Biol.* **1987**, *105*, 755.

(49) Wawrezinieck, L.; Rigneault, H.; Marguet, D.; Lenne, P.-F. Fluorescence Correlation Spectroscopy Diffusion Laws to Probe the Submicron Cell Membrane Organization. *Biophys. J.* **2005**, *89*, 4029–4042.

(50) Ruprecht, V.; Wieser, S.; Marguet, D.; Schütz, G. J. Spot Variation Fluorescence Correlation Spectroscopy Allows for Superresolution Chronoscopy of Confinement Times in Membranes. *Biophys. J.* **2011**, *100*, 2839–2845.

(51) Lasserre, R.; Guo, X.-J.; Conchonaud, F.; Hamon, Y.; Hawchar, O.; Bernard, A.-M.; Soudja, S. M.; Lenne, P.-F.; Rigneault, H.; Olive, D.; *et al.* Raft Nanodomains Contribute to Akt/PKB Plasma Membrane Recruitment and Activation. **2008**, *4*, 538.

(52) Simons, K.; Gerl, M. J. Revitalizing Membrane Rafts: New Tools and Insights. *Nat. Rev. Mol. Cell Biol.* **2010**, *11*, 688–699.

(53) Betzig, E.; Patterson, G. H.; Sougrat, R.; Lindwasser, O. W.; Olenych, S.; Bonifacino, J. S.; Davidson, M. W.; Lippincott-Schwartz, J.; Hess, H. F. Imaging Intracellular Fluorescent Proteins at Nanometer Resolution. *Science* **2006**, *313*, 1642.

(54) Hess, S. T.; Girirajan, T. P. K.; Mason, M. D. Ultra-High Resolution Imaging by Fluorescence Photoactivation Localization Microscopy. *Biophys. J.* **2006**, *91*, 4258–4272.

(55) Rust, M. J.; Bates, M.; Zhuang, X. Sub-Diffraction-Limit Imaging by Stochastic Optical Reconstruction Microscopy (STORM). *Nat Meth* **2006**, *3*, 793–796.

(56) Hell, S. W.; Wichmann, J. Breaking the Diffraction Resolution Limit by Stimulated Emission: Stimulated-Emission-Depletion Fluorescence Microscopy. *Opt. Lett.* **1994**, *19*, 780–782.

(57) Klar, T. A.; Jakobs, S.; Dyba, M.; Egner, A.; Hell, S. W. Fluorescence Microscopy with Diffraction Resolution Barrier Broken by Stimulated Emission. *Proc. Natl. Acad. Sci.* **2000**, *97*, 8206–8210.

(58) Hell, S. W. Far-Field Optical Nanoscopy. *Science* **2007**, *316*, 1153–1158.

(59)  Hwang, J.; Gheber, L. A.; Margolis, L.; Edidin, M. Domains in Cell Plasma Membranes Investigated by near-Field Scanning Optical Microscopy. *Biophys. J.* **1998**, *74*, 2184–2190.

(60)  De Lange, F.; Cambi, A.; Huijbens, R.; de Bakker, B.; Rensen, W.; Garcia-Parajo, M.; van Hulst, N.; Figdor, C. G. Cell Biology beyond the Diffraction Limit: Near-Field Scanning Optical Microscopy. *J. Cell Sci.* **2001**, *114*, 4153–4160.

(61)  van Zanten, T. S.; Gómez, J.; Manzo, C.; Cambi, A.; Buceta, J.; Reigada, R.; Garcia-Parajo, M. F. Direct Mapping of Nanoscale Compositional Connectivity on Intact Cell Membranes. *Proc. Natl. Acad. Sci.* **2010**, *107*, 15437–15442.

(62)  van Zanten, T. S.; Cambi, A.; Garcia-Parajo, M. F. A Nanometer Scale Optical View on the Compartmentalization of Cell Membranes. *Biochim. Biophys. Acta BBA - Biomembr.* **2010**, *1798*, 777–787.

(63)  Manzo, C.; van Zanten, T. S.; Garcia-Parajo, M. F. Nanoscale Fluorescence Correlation Spectroscopy on Intact Living Cell Membranes with NSOM Probes. *Biophys. J.* **2011**, *100*, L8–L10.

(64)  Mueller, V.; Ringemann, C.; Honigmann, A.; Schwarzmann, G.; Medda, R.; Leutenegger, M.; Polyakova, S.; Belov, V. N.; Hell, S. W.; Eggeling, C. STED Nanoscopy Reveals Molecular Details of Cholesterol- and Cytoskeleton-Modulated Lipid Interactions in Living Cells. *Biophys. J.* **2011**, *101*, 1651–1660.

(65)  Honigmann, A.; Mueller, V.; Hell, S. W.; Eggeling, C. STED Microscopy Detects and Quantifies Liquid Phase Separation in Lipid Membranes Using a New Far-Red Emitting Fluorescent Phosphoglycerolipid Analogue. *Faraday Discuss.* **2013**, *161*, 77–89.

(66)  Honigmann, A.; Sadeghi, S.; Keller, J.; Hell, S. W.; Eggeling, C.; Vink, R. A Lipid Bound Actin Meshwork Organizes Liquid Phase Separation in Model Membranes. *Elife* **2014**, *3*, e01671.

(67)  Honigmann, A.; Mueller, V.; Ta, H.; Schoenle, A.; Sezgin, E.; Hell, S. W.; Eggeling, C. Scanning STED-FCS Reveals Spatiotemporal Heterogeneity of Lipid Interaction in the Plasma Membrane of Living Cells. *Nat. Commun.* **2014**, *5*, 5412.

(68)  Vicidomini, G.; Ta, H.; Honigmann, A.; Mueller, V.; Clausen, M. P.; Waithe, D.; Galiani, S.; Sezgin, E.; Diaspro, A.; Hell, S. W.; *et al.* STED-FLCS: An Advanced Tool to Reveal Spatiotemporal Heterogeneity of Molecular Membrane Dynamics. *Nano Lett.* **2015**, *15*, 5912–5918.

(69) Sarangi, N. K.; Ayappa, K. G.; Basu, J. K. Complex Dynamics at the Nanoscale in Simple Biomembranes. *Sci. Rep.* **2017**, *7*, 11173.

(70) Manzo, C.; Garcia-Parajo, M. F. A Review of Progress in Single Particle Tracking: From Methods to Biophysical Insights. *Rep. Prog. Phys.* **2015**, *78*, 124601.

(71) Ortega-Arroyo, J.; Kukura, P. Interferometric Scattering Microscopy (iSCAT): New Frontiers in Ultrafast and Ultrasensitive Optical Microscopy. *Phys. Chem. Chem. Phys.* **2012**, *14*, 15625.

(72) Spillane, K. M.; Ortega-Arroyo, J.; de Wit, G.; Eggeling, C.; Ewers, H.; Wallace, M. I.; Kukura, P. High-Speed Single-Particle Tracking of GM1 in Model Membranes Reveals Anomalous Diffusion due to Interleaflet Coupling and Molecular Pinning. *Nano Lett.* **2014**, *14*, 5390–5397.

(73) Spindler, S.; Ehrig, J.; König, K.; Nowak, T.; Piliarik, M.; Stein, H. E.; Taylor, R. W.; Garanger, E.; Lecommandoux, S.; Alves, I. D.; *et al.* Visualization of Lipids and Proteins at High Spatial and Temporal Resolution via Interferometric Scattering (iSCAT) Microscopy. *J. Phys. Appl. Phys.* **2016**, *49*, 274002.

(74) Wu, H.-M.; Lin, Y.-H.; Yen, T.-C.; Hsieh, C.-L. Nanoscopic Substructures of Raft-Mimetic Liquid-Ordered Membrane Domains Revealed by High-Speed Single-Particle Tracking. *Sci. Rep.* **2016**, *6*, 20542.

(75) Holzmeister, P.; Acuna, G. P.; Grohmann, D.; Tinnefeld, P. Breaking the Concentration Limit of Optical Single-Molecule Detection. *Chem Soc Rev* **2014**, *43*, 1014–1028.

(76) Punj, D.; Ghenuche, P.; Moparthi, S. B.; de Torres, J.; Grigoriev, V.; Rigneault, H.; Wenger, J. Plasmonic Antennas and Zero-Mode Waveguides to Enhance Single Molecule Fluorescence Detection and Fluorescence Correlation Spectroscopy toward Physiological Concentrations. *Wiley Interdiscip. Rev. Nanomed. Nanobiotechnol.* **2014**, *6*, 268–282.

(77) Wenger, J.; Rigneault, H. Photonic Methods to Enhance Fluorescence Correlation Spectroscopy and Single Molecule Fluorescence Detection. *Int. J. Mol. Sci.* **2010**, *11*, 206–221.

(78) Levene, M. J.; Korlach, J.; Turner, S. W.; Foquet, M.; Craighead, H. G.; Webb, W. W. Zero-Mode Waveguides for Single-Molecule Analysis at High Concentrations. *Science* **2003**, *299*, 682.

(79) Genet, C.; Ebbesen, T. W. Light in Tiny Holes. *Nature* **2007**, *445*, 39–46.

(80) Moran-Mirabal, J. M.; Craighead, H. G. Zero-Mode Waveguides: Sub-Wavelength Nanostructures for Single Molecule Studies at High Concentrations. *Single Mol. Detect. Ther. Technol.* **2008**, *46*, 11–17.

(81) Samiee, K. T.; Moran-Mirabal, J. M.; Cheung, Y. K.; Craighead, H. G. Zero Mode Waveguides for Single-Molecule Spectroscopy on Lipid Membranes. *Biophys. J.* **2006**, *90*, 3288–3299.

(82) Wenger, J.; Rigneault, H.; Dintinger, J.; Marguet, D.; Lenne, P.-F. Single-Fluorophore Diffusion in a Lipid Membrane over a Subwavelength Aperture. *J. Biol. Phys.* **2006**, *32*, SN1-SN4.

(83) Edel, J. B.; Wu, M.; Baird, B.; Craighead, H. G. High Spatial Resolution Observation of Single-Molecule Dynamics in Living Cell Membranes. *Biophys. J.* **2005**, *88*, L43–L45.

(84) Jose M Moran-Mirabal and Alexis J Torres and Kevan T Samiee and Barbara A Baird and Harold G Craighead. Cell Investigation of Nanostructures: Zero-Mode Waveguides for Plasma Membrane Studies with Single Molecule Resolution. *Nanotechnology* **2007**, *18*, 195101.

(85) Richards, C. I.; Luong, K.; Srinivasan, R.; Turner, S. W.; Dougherty, D. A.; Korlach, J.; Lester, H. A. Live-Cell Imaging of Single Receptor Composition Using Zero-Mode Waveguide Nanostructures. *Nano Lett.* **2012**, *12*, 3690–3694.

(86) Wenger, J.; Conchonaud, F.; Dintinger, J.; Wawrezinieck, L.; Ebbesen, T. W.; Rigneault, H.; Marguet, D.; Lenne, P.-F. Diffusion Analysis within Single Nanometric Apertures Reveals the Ultrafine Cell Membrane Organization. *Biophys. J.* **2007**, *92*, 913–919.

(87) Gérard, D.; Wenger, J.; Bonod, N.; Popov, E.; Rigneault, H.; Mahdavi, F.; Blair, S.; Dintinger, J.; Ebbesen, T. W. Nanoaperture-Enhanced Fluorescence: Towards Higher Detection Rates with Plasmonic Metals. *Phys. Rev. B* **2008**, *77*, 045413.

(88) Kelly, C. V.; Baird, B. A.; Craighead, H. G. An Array of Planar Apertures for Near-Field Fluorescence Correlation Spectroscopy. *Biophys. J.* **2011**, *100*, L34–L36.

(89) Kelly, C. V.; Wakefield, D. L.; Holowka, D. A.; Craighead, H. G.; Baird, B. A. Near-Field Fluorescence Cross-Correlation Spectroscopy on Planar Membranes. *ACS Nano* **2014**, *8*, 7392–7404.

(90) Schuller, J. A.; Barnard, E. S.; Cai, W.; Jun, Y. C.; White, J. S.; Brongersma, M. L. Plasmonics for Extreme Light Concentration and Manipulation. *Nat Mater* **2010**, *9*, 193–204.

(91) Novotny, L.; van Hulst, N. Antennas for Light. *Nat. Photonics* **2011**, *5*, 83–90.

(92) Biagioni, P.; Huang, J.-S.; Hecht, B. Nanoantennas for Visible and Infrared Radiation. *Rep. Prog. Phys.* **2012**, *75*, 024402.

(93) Halas, N. J.; Lal, S.; Chang, W.-S.; Link, S.; Nordlander, P. Plasmons in Strongly Coupled Metallic Nanostructures. *Chem. Rev.* **2011**, *111*, 3913–3961.

(94) Koenderink, A. F. Single-Photon Nanoantennas. *ACS Photonics* **2017**, *4*, 710–722.

(95) Kinkhabwala, A.; Yu, Z.; Fan, S.; Avlasevich, Y.; Müllen, K.; Moerner, W. E. Large Single-Molecule Fluorescence Enhancements Produced by a Bowtie Nanoantenna. *Nat. Photonics* **2009**, *3*, 654–657.

(96) Acuna, G. P.; Moller, F. M.; Holzmeister, P.; Beater, S.; Lalkens, B.; Tinnefeld, P. Fluorescence Enhancement at Docking Sites of DNA-Directed Self-Assembled Nanoantennas. *Science* **2012**, *338*, 506–510.

(97) Puchkova, A.; Vietz, C.; Pibiri, E.; Wünsch, B.; Sanz Paz, M.; Acuna, G. P.; Tinnefeld, P. DNA Origami Nanoantennas with over 5000-Fold Fluorescence Enhancement and Single-Molecule Detection at 25 μM. *Nano Lett.* **2015**, *15*, 8354–8359.

(98) Yuan, H.; Khatua, S.; Zijlstra, P.; Yorulmaz, M.; Orrit, M. Thousand-Fold Enhancement of Single-Molecule Fluorescence Near a Single Gold Nanorod. *Angew. Chem. Int. Ed.* **2013**, *52*, 1217–1221.

(99) Khatua, S.; Paulo, P. M. R.; Yuan, H.; Gupta, A.; Zijlstra, P.; Orrit, M. Resonant Plasmonic Enhancement of Single-Molecule Fluorescence by Individual Gold Nanorods. *ACS Nano* **2014**, *8*, 4440–4449.

(100) Lohmüller, T.; Iversen, L.; Schmidt, M.; Rhodes, C.; Tu, H.-L.; Lin, W.-C.; Groves, J. T. Single Molecule Tracking on Supported Membranes with Arrays of Optical Nanoantennas. *Nano Lett.* **2012**, *12*, 1717–1721.

(101) Flauraud, V.; van Zanten, T. S.; Mivelle, M.; Manzo, C.; Garcia Parajo, M. F.; Brugger, J. Large-Scale Arrays of Bowtie Nanoaperture Antennas for Nanoscale Dynamics in Living Cell Membranes. *Nano Lett.* **2015**, *15*, 4176–4182.

(102) Flynn, J. D.; Haas, B. L.; Biteen, J. S. Plasmon-Enhanced Fluorescence from Single Proteins in Living Bacteria. *J. Phys. Chem. C* **2016**, *120*, 20512–20517.

(103) Fabrizio, E. D.; Schlücker, S.; Wenger, J.; Regmi, R.; Rigneault, H.; Calafiore, G.; West, M.; Cabrini, S.; Fleischer, M.; van Hulst, N. F.; *et al.* Roadmap on Biosensing and Photonics with Advanced Nano-Optical Methods. *J. Opt.* **2016**, *18*, 063003.

(104) Perozziello, G.; Candeloro, P.; De Grazia, A.; Esposito, F.; Allione, M.; Coluccio, M. L.; Tallerico, R.; Valpapuram, I.; Tirinato, L.; Das, G.; *et al.* Microfluidic Device for Continuous Single Cells Analysis via Raman Spectroscopy Enhanced by Integrated Plasmonic Nanodimers. *Opt. Express* **2016**, *24*, A180–A190.

(105) Coluccio, M. L.; Gentile, F.; Das, G.; Nicastri, A.; Perri, A. M.; Candeloro, P.; Perozziello, G.; Proietti Zaccaria, R.; Gongora, J. S. T.; Alrasheed, S.; *et al.* Detection of Single Amino Acid Mutation in Human Breast Cancer by Disordered Plasmonic Self-Similar Chain. *Sci. Adv.* **2015**, *1*, e1500487.

(106) Langguth, L.; Femius Koenderink, A. Simple Model for Plasmon Enhanced Fluorescence Correlation Spectroscopy. *Opt. Express* **2014**, *22*, 15397–15409.

(107) Estrada, L. C.; Aramendía, P. F.; Martínez, O. E. 10000 Times Volume Reduction for Fluorescence Correlation Spectroscopy Using Nano-Antennas. *Opt. Express* **2008**, *16*, 20597–20602.

(108) Wang, Q.; Lu, G.; Hou, L.; Zhang, T.; Luo, C.; Yang, H.; Barbillon, G.; Lei, F. H.; Marquette, C. A.; Perriat, P.; *et al.* Fluorescence Correlation Spectroscopy near Individual Gold Nanoparticle. *Chem. Phys. Lett.* **2011**, *503*, 256–261.

(109) Lu, G.; Liu, J.; Zhang, T.; Li, W.; Hou, L.; Luo, C.; Lei, F.; Manfait, M.; Gong, Q. Plasmonic near-Field in the Vicinity of a Single Gold Nanoparticle Investigated with Fluorescence Correlation Spectroscopy. *Nanoscale* **2012**, *4*, 3359–3364.

(110) Punj, D.; de Torres, J.; Rigneault, H.; Wenger, J. Gold Nanoparticles for Enhanced Single Molecule Fluorescence Analysis at Micromolar Concentration. *Opt. Express* **2013**, *21*, 27338.

(111) Khatua, S.; Yuan, H.; Orrit, M. Enhanced-Fluorescence Correlation Spectroscopy at Micro-Molar Dye Concentration around a Single Gold Nanorod. *Phys Chem Chem Phys* **2015**, *17*, 21127–21132.

(112) Kinkhabwala, A. A.; Yu, Z.; Fan, S.; Moerner, W. E. Fluorescence Correlation Spectroscopy at High Concentrations Using Gold Bowtie Nanoantennas. *Single Mol. Spectrosc. Curr. Status Perspect.* **2012**, *406*, 3–8.

(113) Dutta Choudhury, S.; Ray, K.; Lakowicz, J. R. Silver Nanostructures for Fluorescence Correlation Spectroscopy: Reduced Volumes and Increased Signal Intensities. *J. Phys. Chem. Lett.* **2012**, *3*, 2915–2919.

(114) Punj, D.; Regmi, R.; Devilez, A.; Plauchu, R.; Moparthi, S. B.; Stout, B.; Bonod, N.; Rigneault, H.; Wenger, J. Self-Assembled Nanoparticle Dimer Antennas for

(114) Plasmonic-Enhanced Single-Molecule Fluorescence Detection at Micromolar Concentrations. *ACS Photonics* **2015**, *2*, 1099–1107.

(115) Pradhan, B.; Khatua, S.; Gupta, A.; Aartsma, T.; Canters, G.; Orrit, M. Gold-Nanorod-Enhanced Fluorescence Correlation Spectroscopy of Fluorophores with High Quantum Yield in Lipid Bilayers. *J. Phys. Chem. C* **2016**, *120*, 25996–26003.

(116) Punj, D.; Mivelle, M.; Moparthi, S. B.; van Zanten, T. S.; Rigneault, H.; van Hulst, N. F.; García-Parajó, M. F.; Wenger, J. A Plasmonic "antenna-in-Box" Platform for Enhanced Single-Molecule Analysis at Micromolar Concentrations. *Nat. Nanotechnol.* **2013**, *8*, 512–516.

(117) Ghenuche, P.; de Torres, J.; Moparthi, S. B.; Grigoriev, V.; Wenger, J. Nanophotonic Enhancement of the Förster Resonance Energy-Transfer Rate with Single Nanoapertures. *Nano Lett.* **2014**, *14*, 4707–4714.

(118) Flauraud, V.; Regmi, R.; Winkler, P. M.; Alexander, D. T. L.; Rigneault, H.; van Hulst, N. F.; García-Parajo, M. F.; Wenger, J.; Brugger, J. In-Plane Plasmonic Antenna Arrays with Surface Nanogaps for Giant Fluorescence Enhancement. *Nano Lett.* **2017**, 1703–1710.

(119) Winkler, P. M.; Regmi, R.; Flauraud, V.; Brugger, J.; Rigneault, H.; Wenger, J.; García-Parajo, M. F. Transient Nanoscopic Phase Separation in Biological Lipid Membranes Resolved by Planar Plasmonic Antennas. *ACS Nano* **2017**, *11*, 7241–7250.

(120) Brown, D. A. Seeing Is Believing: Visualization of Rafts in Model Membranes. *Proc. Natl. Acad. Sci.* **2001**, *98*, 10517–10518.

(121) Bagatolli, L. A.; Gratton, E. Two Photon Fluorescence Microscopy of Coexisting Lipid Domains in Giant Unilamellar Vesicles of Binary Phospholipid Mixtures. *Biophys. J.* **2000**, *78*, 290–305.

(122) Yasuda, T.; Tsuchikawa, H.; Murata, M.; Matsumori, N. Deuterium NMR of Raft Model Membranes Reveals Domain-Specific Order Profiles and Compositional Distribution. *Biophys. J.* **2015**, *108*, 2502–2506.

(123) Apajalahti, T.; Niemela, P.; Govindan, P. N.; Miettinen, M. S.; Salonen, E.; Marrink, S.-J.; Vattulainen, I. Concerted Diffusion of Lipids in Raft-like Membranes. *Faraday Discuss.* **2010**, *144*, 411–430.

(124) Sodt, A. J.; Sandar, M. L.; Gawrisch, K.; Pastor, R. W.; Lyman, E. The Molecular Structure of the Liquid-Ordered Phase of Lipid Bilayers. *J. Am. Chem. Soc.* **2014**, *136*, 725–732.

(125) Sodt, A. J.; Pastor, R. W.; Lyman, E. Hexagonal Substructure and Hydrogen Bonding in Liquid-Ordered Phases Containing Palmitoyl Sphingomyelin. *Biophys. J.* **2015**, *109*, 948–955.

(126) Silvius, J. R. Fluorescence Energy Transfer Reveals Microdomain Formation at Physiological Temperatures in Lipid Mixtures Modeling the Outer Leaflet of the Plasma Membrane. *Biophys. J.* **2003**, *85*, 1034–1045.

(127) de Almeida, R. F. M.; Loura, L. M. S.; Fedorov, A.; Prieto, M. Lipid Rafts Have Different Sizes Depending on Membrane Composition: A Time-Resolved Fluorescence Resonance Energy Transfer Study. *J. Mol. Biol.* **2005**, *346*, 1109–1120.

(128) Regmi, R.; Winkler, P. M.; Flauraud, V.; Borgman, K. J. E.; Manzo, C.; Brugger, J.; Rigneault, H.; Wenger, J.; García-Parajo, M. F. Planar Optical Nanoantennas Resolve Cholesterol-Dependent Nanoscale Heterogeneities in the Plasma Membrane of Living Cells. *Nano Lett.* **2017**, *17*, 6295–6302.

(129) Nicolau, D. V.; Burrage, K.; Parton, R. G.; Hancock, J. F. Identifying Optimal Lipid Raft Characteristics Required To Promote Nanoscale Protein-Protein Interactions on the Plasma Membrane. *Mol. Cell. Biol.* **2006**, *26*, 313–323.